\documentclass[%
 reprint,
 amsmath,amssymb,
 aps,
]{revtex4-2}
\usepackage{subfigure}
\usepackage{overpic}
\usepackage{graphicx}
\usepackage{dcolumn}
\usepackage{bm}
\usepackage[colorlinks,linkcolor=blue,citecolor=blue]{hyperref}
\usepackage{lineno}
\usepackage{graphicx}

\usepackage{url}
\usepackage{ulem}
\usepackage{amsmath}
\usepackage{amsfonts}
\usepackage{textcomp}
\usepackage{subfigure}
\usepackage{verbatim}
\usepackage{rotating}
\usepackage{tikz}
\usepackage{graphicx}
\usepackage{color}
\usepackage{epstopdf}
\def \ee   {e^+e^-}

\begin{document}

\title{\bf \boldmath Search for the Rare Decays $D_s^+\to h^+(h^{0})e^+e^-$}

\author{
M.~Ablikim$^{1}$, M.~N.~Achasov$^{4,b}$, P.~Adlarson$^{75}$, O.~Afedulidis$^{3}$, X.~C.~Ai$^{80}$, R.~Aliberti$^{35}$, A.~Amoroso$^{74A,74C}$, Q.~An$^{71,58}$, Y.~Bai$^{57}$, O.~Bakina$^{36}$
% ...
\\
\vspace{0.2cm}
(BESIII Collaboration)\\
\vspace{0.2cm} 
\it
$^{1}$ Institute of High Energy Physics, Beijing 100049, People's Republic of China\\
% ...
$^{3}$ Bochum  Ruhr-University, D-44780 Bochum, Germany\\
$^{4}$ Budker Institute of Nuclear Physics SB RAS (BINP), Novosibirsk 630090, Russia\\
% ...
$^{35}$ Johannes Gutenberg University of Mainz, Johann-Joachim-Becher-Weg 45, D-55099 Mainz, Germany\\
$^{36}$ Joint Institute for Nuclear Research, 141980 Dubna, Moscow region, Russia\\
% ...
$^{57}$ Southeast University, Nanjing 211100, People's Republic of China\\
$^{58}$ State Key Laboratory of Particle Detection and Electronics, Beijing 100049, Hefei 230026, People's Republic of China\\
% ...
$^{71}$ University of Science and Technology of China, Hefei 230026, People's Republic of China\\
% ...
$^{74}$ University of Turin and INFN, (A)University of Turin, I-10125, Turin, Italy; (B)University of Eastern Piedmont, I-15121, Alessandria, Italy; (C)INFN, I-10125, Turin, Italy\\
$^{75}$ Uppsala University, Box 516, SE-75120 Uppsala, Sweden\\
% ...
$^{80}$ Zhengzhou University, Zhengzhou 450001, People's Republic of China\\
\vspace{0.2cm}
% ...
$^{b}$ Also at the Novosibirsk State University, Novosibirsk, 630090, Russia\\
% ...
\vspace{0.4cm}
}


\begin{abstract}


Using 7.33~fb$^{-1}$ of $\ee$ collision data collected by the BESIII detector at center-of-mass energies in the range of $\sqrt{s}=4.128 - 4.226$~GeV, we search for the rare decays $D_{s}^+\to h^+(h^{0})e^{+}e^{-}$, where $h$ represents a kaon or pion. 
By requiring the $\ee$ invariant mass to be consistent with a $\phi(1020)$, 
$0.98<M(e^{+}e^{-})<1.04$ ~GeV/$c^2$, 
the decay $D_s^+\to\pi^+\phi,\phi\to\ee$ is observed with a statistical significance of 7.8$\sigma$, and evidence for the decay $D_s^+\to\rho^+\phi,\phi\to\ee$ is found for the first time with a statistical significance of 4.4$\sigma$. 
The decay branching fractions are measured to be $\mathcal{B}(D_s^+\to\pi^+\phi, \phi\to\ee )=(1.17^{+0.23}_{-0.21}\pm0.03)\times 10^{-5}$, and 
 $\mathcal{B}(D_s^+\to\rho^+\phi, \phi\to\ee )=(2.44^{+0.67}_{-0.62}\pm 0.16)\times 10^{-5}$,  where the first uncertainties are statistical and the second systematic. No significant signal for the three four-body decays of $D_{s}^{+}\to \pi^{+}\pi^{0}\ee,\ D_{s}^{+}\to K^{+}\pi^{0}\ee$, and $D_{s}^{+}\to K_{S}^{0}\pi^{+}\ee$ is observed. For $D_{s}^{+}\to \pi^{+}\pi^{0}\ee$, the $\phi$ mass region is vetoed to minimize the long-distance effects. The 90$\%$ confidence level upper limits set on the branching fractions of these decays are in the range of $(7.0-8.1)\times 10^{-5}$.

\end{abstract}

\oddsidemargin  -0.2cm
\evensidemargin -0.2cm
\maketitle

In flavor physics, rare decays play an important role in precision tests of the Standard Model (SM) and as probes of physics beyond the SM. 
In the SM, the rare decays $D_{s}^+\to h^{+}(h^{0})e^{+}e^{-}$, where $h$ denotes a pion or kaon, involve both short-distance (SD) and long-distance (LD) contributions. The SD contributions proceed via the 
$c\to u l^+l^-$ flavor-changing neutral-current (FCNC) transition, which can only occur at the loop level and are highly suppressed by the Glashow$-$Iliopoulos$-$Maiani (GIM) mechanism~\cite{Glashow:1970gm}. 
The GIM suppression is more effective in the charm sector compared to the bottom and strange sectors, leading to naive SD-only branching fractions (BFs) as low as $10^{-9}$~\cite{Paul:2011ar,deBoer:2015boa, Burdman:2001tf, Paul:2011ar}.
The tiny SM contribution makes the FCNC transitions in the charm sector particularly sensitive to new physics, which may significantly enhance the BFs through
the presence of new particles and interactions~\cite{Burdman:2001tf, Fajfer:2015mia, Paul:2011ar,deBoer:2015boa}.
The LD contributions, occurring through a radiated photon or an intermediate meson decaying to dileptons, dominate the decays of $D_{s}^+ \to h^+(h^{0})e^{+}e^{-}$ and can enhance the BFs to the order of $10^{-6}$~\cite{Paul:2012ab, Cappiello:2012vg, Sanchez:2022vyq}. Furthermore, as $D_s^+ \to V e^+e^-$ decays ($V$ is a light vector meson) receive considerable contributions from virtual photons, and $D_s^+ \to V \gamma$ decays are predicted to have BFs as high as ${\cal O}(10^{-3})$~\cite{deBoer:2017que}, one would expect  the BFs of $D_s^+ \to V \ee$ to reach $10^{-5}$. 
Therefore, the FCNC processes are often overshadowed by the LD effects. The SD effects can be accessed through measurements in the dilepton mass regions away from those of the intermediate mesons, such as $\eta,\rho,\omega$ 
and $\phi$. 
Moreover, measurements of angular dependence and the charge-parity-conjugation asymmetry in regions away from or dominated by the resonances as SM null tests are also helpful for exploring the SD effects~\cite{Burdman:2001tf, Fajfer:2015mia, Paul:2011ar,deBoer:2015boa, Paul:2012ab, Cappiello:2012vg, Sanchez:2022vyq,Gisbert:2020vjx}. 

Recently, experimental results for the tests of lepton universality (LU) in the beauty-quark FCNC decays $b\to sl^+l^-$ have received significant attention~\cite{LHCb:2021trn}. The latest measurement from LHCb shows the tests of muon-electron universality using $B^+\to K^+l^+l^-$ and $B^0\to K^*l^+l^-$ decays are in agreement with the SM predictions~\cite{LHCb:2022qnv}. An analogous measurement of $c\to ul^+l^-$ provides an important complementary test of lepton universality in the up-quark sector~\cite{Ke:2023qzc}.
However due to limited experimental results on di-electron modes, LU tests in $c\to u l^+ l^-$ are largely unexplored~\cite{Gisbert:2020vjx}.
These modes also provide sensitivity to a broader class of new physics, 
which could affect angular observables and CP asymmetries at a few percent level in $D^0\to h^+h^- l^+ l^-$ decays~\cite{DeBoer:2018pdx}.

Experimentally, the decays $D^0\to K^-\pi^+\mu^+\mu^-$, $D^0\to \pi^+\pi^-\mu^+\mu^-$, $D^0\to K^-K^+\mu^+\mu^-$, and $D^0\to K^-\pi^+\ee$ have been observed with BFs of ${\cal O}(10^{-6}-10^{-7})$ in the LHCb and BaBar experiments~\cite{LHCb:2015yuk, LHCb:2017uns, BaBar:2018bwm}, where the decays are dominated by the LD contributions with lepton pairs originating from $\rho$ and $\omega$ resonances.  Lately, LHCb experiment reports the measurement of the branching fraction ratio $\mathcal{B}(\phi\to\mu^+\mu^-)/\mathcal{B}(\phi\to e^+e^-)$ with $D_{(s)}^+\to \pi^+\phi$ decays, and the result is compatible with the SM predictions~\cite{LHCb:2024vdc}.
Additionally, the FCNC decays of $D^{0,\pm}$ mesons have been searched for extensively~\cite{BESIII:2018hqu,LHCb:2020car}, but the present  experimental upper bounds are still much higher than the SM predictions~\cite{HeavyFlavorAveragingGroup:2022wzx}.
For the rare decays of $D_s^+$ meson, the upper limits on the three-body decays $D_{s}^+\to h^+ l^{+}l^{-}$, where the dilepton mass is away from the $\phi$ mass, are in the range ${\cal O}(10^{-8}-10^{-6})$ as reported in the BaBar~\cite{BaBar:2011ouc} and LHCb experiments~\cite{LHCb:2020car}.  

In this Letter, we measure the LD contributions of $D_{s}^+\to\pi^+\phi$ with $\phi\to\ee$, $D_{s}^+\to\rho^+\phi$ with $\rho^+\to\pi^+\pi^0$ and $\phi\to\ee$, and search for the four-body decays of $D_s^+\to\pi^{+}\pi^{0}\ee$, $D_s^+\to K^{+}\pi^{0}e^{+}e^{-}$, $D_s^+\to K_S^{0}\pi^{+}e^{+}e^{-}$ using data samples corresponding to an integrated luminosity of $7.33$~fb$^{-1}$ accumulated with the BESIII detector at $\ee$ center-of-mass (c.m.) energies in the range $\sqrt{s}=4.128 - 4.226$~GeV~\cite{BESIII:2020eyu,BESIII:2022dxl}. In these data samples, the $D_s^{\pm}$ mesons are dominantly produced in the process $\ee\to D_s^{*\pm} D_s^{\mp}$ with $D_s^{*\pm}$ predominantly decaying via $D_s^{*\pm}\to D_s^{\pm}\gamma$~\cite{CLEO:2008ojp}. 
The cross section of $\ee\to D_s^{*\pm} D_s^{\mp}$ is about 20 times larger than that of $\ee\to D_s^{+} D_s^{-}$.
Consequently, the $D_{s}^+$ candidates are selected from the process of $\ee\to D_s^{*\pm}(\to D_s^{\pm}\gamma) D_s^{\mp}$ in this work. 
Throughout this Letter, charged conjugate modes are included, and $\rho$ denotes the $\rho(770)$.

Details about the BESIII detector are described in Refs.~\cite{BESIII:2009fln,Huang:2022wuo}.
Simulated Monte-Carlo (MC) event samples produced with a {\sc
geant4}-based~\cite{geant4} package, which
includes the geometric description of the BESIII detector and the
detector response, are used to determine detection efficiencies
and to estimate backgrounds. 
The simulation models the beam energy spread and initial state radiation (ISR) in the $e^+e^-$ annihilations with the generator {\sc kkmc}~\cite{ref:kkmc}.
The input cross-section line shape of $\ee\to D_s^{*\pm}D_s^{\mp}$ is modelled according to the measurement in Ref.~\cite{ref:LS}.
The signal $D_s^+\to\pi^{+}\phi$ ($D_s^+\to\rho^{+}\phi$) is generated using scalar-to-vector-scalar (scalar-to-two-vectors) model in {\sc
evtgen}~\cite{ref:evtgen}. For the $\phi\to\ee$ ($\rho^+\to\pi^+\pi^0$) decay, the vector-to-a-lepton-pair (vector-to-two-scalars) model is employed.
The signal processes of $D_s^+\to\pi^{+}\pi^{0}e^{+}e^{-}$,~$K^{+}\pi^{0}e^{+}e^{-}$, and  $K_{s}^{0}\pi^{+}e^{+}e^{-}$ are generated uniformly in phase space.
The inclusive MC sample, corresponding to 40 times the integrated luminosity of data, includes the production of open charm
processes, the ISR production of vector charmonium(-like) states,
and the continuum processes incorporated in {\sc
kkmc}.
All particle decays are modelled with {\sc
evtgen} using BFs
either taken from the
Particle Data Group (PDG)~\cite{pdg} when available,
or otherwise estimated with {\sc lundcharm}~\cite{ref:lundcharm}.
Final state radiation
from charged final state particles is incorporated using the {\sc
photos} package~\cite{photos}.

We apply a single tag (ST) method to search for $D_s^+$ candidates.
The ST method requires one $D_s^{+}$ meson to be fully reconstructed
in the signal mode in each event.
The BF of $D_s^+\to h^+(h^0)\ee$ is given by
\begin{equation}
\mathcal{B}(D_s^+\to h^{+}(h^{0})\ee)=\frac{N_{\rm{sig}}}{2\cdot N_{D_s^{*\pm} D_s^{\mp}} \cdot \epsilon \cdot \mathcal{B}_{\rm{inter}}},
\label{eq:bf}
\end{equation}
where $N_{\rm{sig}}$ is the signal yield and $N_{D_s^{*\pm} D_s^{\mp}}=(64.72\pm 0.28)\times 10^{5}$ is the total number of $D_s^{*\pm} D_s^{\mp}$ pairs in the data samples~\cite{ref:tot_DD}.
The signal efficiency $\epsilon$ weighted over eight energy points is given by
$\epsilon=\sum^i \epsilon^i N_{D_s^{*\pm} D_s^{\mp}}^i/N_{D_s^{*\pm} D_s^{\mp}}$, where $\epsilon^i$ and $N_{D_s^{*\pm} D_s^{\mp}}^i$ are the detection efficiency and the number of $D_s^{*\pm} D_s^{\mp}$ pairs at the $i$-th energy point, respectively, and $\mathcal{B}_{\rm{inter}}$ is the product BF of intermediate state decays.

The final states include $\pi^+$, $\pi^0$, $K^+$, $K_S^0$, and $e^\pm$ particles.  We require that charged tracks detected in the main drift chamber (MDC) satisfy $|\rm{cos\theta}|<0.93$, where the polar angle $\theta$ is defined with respect to the $z$-axis, which is the symmetry axis of the MDC. For charged tracks not used for reconstruction of $K_S^0$, the distance of the closest approach to the interaction point (IP) 
must be less than 10\,cm along the $z$-axis, $|V_{z}|$, and less than 1\,cm in the transverse plane, $|V_{xy}|$.
Particle identification~(PID) is applied for charged pion and kaon tracks by combining the measurements of the specific ionization energy loss in the MDC~(d$E$/d$x$) and the flight time in the time-of-flight system (TOF) to form likelihoods $\mathcal{L}(h)~(h=K,\pi)$ for each hadron hypothesis. 
Kaons are identified by requiring $\mathcal{L}(K)>\mathcal{L}(\pi)$, while pions are identified by requiring $\mathcal{L}(K)<\mathcal{L}(\pi)$. To suppress the contamination from $K_S^0\to\pi^+\pi^-$ decays, the vertices of all $\pi^+\pi^-$ combinations are required to be less than three times the vertex resolution from the IP. Electron PID uses the measured information in the MDC, TOF, and electromagnetic calorimeter (EMC). Electron candidates are required to satisfy $\mathcal{L}'(e)>0.001$ and $\mathcal{L}'(e)/(\mathcal{L}'(e)+\mathcal{L}'(\pi)+\mathcal{L}'(K))>0.8$. To reduce background
from hadrons and muons, the electron and positron candidates are required to have deposited energy in the EMC greater than 0.8 (0.7) times its momentum for those with momentum larger (less) than 0.4~GeV$/c$~\cite{BESIII:2021duu}. 
To effectively suppress electron-positron pairs originating from $\gamma$-conversions, events with the distance from the reconstructed vertex point of the $\ee$ pair to the IP in the range (2.0, 8.0)~cm are discarded. 
Furthermore, any photons within a cone of $5^\circ$ around the electron direction is recovered to the electron momentum to improve the electron momentum resolution. 

The $K_{S}^0$ candidates are reconstructed from two oppositely charged tracks satisfying $|V_{z}|<$ 20~cm. The two charged tracks are assigned as $\pi^+\pi^-$ without imposing further PID criteria. They are constrained to originate from a common vertex and are required to have an invariant mass within $|M(\pi^{+}\pi^{-}) - m_{K_{S}^{0}}|<$ 12~MeV$/c^{2}$, where $m_{K_{S}^{0}}$ is the $K^0_{S}$ nominal mass~\cite{pdg}. The decay length of the $K^0_S$ candidate must be greater than twice the vertex resolution. The $\pi^0$ candidate is reconstructed from a pair of photon candidates, which are reconstructed using showers in the EMC detector. The showers are required to have  energy deposited greater than 25~MeV in the barrel region ($|\cos \theta|< 0.80$) and 50~MeV in the end cap region ($0.86 <|\cos \theta|< 0.92$). 
The difference between the EMC time and the event start time for each photon is required to be within [0, 700]\,ns to suppress electronic noise and showers unrelated to the event.
The angle between the vectors from the IP to the EMC cluster position and the projection of the closest charged track  at the EMC 
for each photon is required to be
greater than $10^\circ$. The $\pi^0$ candidates must have a $\gamma\gamma$ invariant mass in the region $[0.115, 0.150]$~GeV/c$^2$. 
The backgrounds associated with mis-paired photons are effectively suppressed by requiring the energy of $\pi^0$ candidate to be greater than 0.17~GeV. A kinematic fit constraining the $\gamma\gamma$ invariant mass to the $\pi^0$ nominal mass is performed to improve the $\pi^0$ four-vector for use in later kinematic calculations.  

Candidate $D_{s}^+$ mesons are formed from combinations of identified hadrons and lepton pairs. 
The $h^+(h^{0})\ee$ invariant mass is required to be within $[1.88, 2.02]$ GeV$/c^2$.
To further suppress backgrounds and identify the $D_{s}^+$ candidates from $\ee\to D_s^{*\pm} D_s^{\mp}$ process,  
we introduce two variables, the recoil mass of $D_s^+$, $M_{\rm rec}$, and the mass difference, $\Delta M$:
\begin{equation}
\begin{split}
 &M_{\rm rec}=\sqrt{\left( E_{\rm cm}-\sqrt{|\vec{P}_{D_s^+}|^2+m_{D_s^+}^2}\right)^2-|P_{D_s^+}|^2},\\
 &\Delta M=M(D_s^+\gamma)-M(D_s^+),\\
 \end{split}
\end{equation}
where $E_{\rm cm}$ is the c.m.~energy of $\ee$ system, $\vec{P}_{D_s^+}$ is the three-momentum of the $D_s^+$ in the $\ee$ c.m.~frame, and $m_{D_s^+}$ is the $D_s^+$ nominal mass~\cite{pdg}. $M(D_s^+\gamma)$ is the invariant mass of the $D_s^+$ candidate and the photon candidate for the $D_s^{*+}\to\gamma D_s^+$ process.  If there are multiple photon candidates, the one with the recoil mass of the $D_s^+\gamma$ closest to $m_{D_s^+}$ is chosen.
$M(D_s^+)$ is the invariant mass of the $D_s^+$ candidate.
The signals peak in the $M_{\rm rec}$ spectrum around $ m_{D_s^{*-}}$ for a $D_s^+$ from the $\ee\to D_s^{+}D_s^{*-}$ process or in the $\Delta M$ spectrum around $m_{D_s^{*+}}-m_{D_s^{+}}$ for a $D_s^+$ from $\ee\to D_s^{-}D_s^{*+}(D_s^{*+}\to\gamma D_s^{+})$ process~\cite{BESIII:2020ctr}.
Signal candidates are required to lie within the signal windows in the two-dimensional plane of $M_{\rm rec}$ versus $\Delta M$ listed in the Supplemental Material~\cite{supp_mat}. 
These windows are determined independently for each energy point and decay mode based on the figure-of-merit of $S/\sqrt{S+B}$ for the two $\phi$ decay modes or $\epsilon/(1.5+\sqrt{B})$~\cite{Punzi:2003bu} for the three four-body decays. Two different figure-of-merit are utilized due to the significantly different contributions of the SM in the two cases.
Here, $S$ represents the expected yield from the SM contribution, while $B$ denotes the background yield from the inclusive MC sample.

For the two $\phi$ decay modes, the invariant mass of the electron pair, $M(\ee)$ (as shown in the Supplemental Material~\cite{supp_mat}),  must lie within the $\phi$ mass window of $[0.98,1.04]$~GeV$/c^2$, determined from the signal MC simulation. 
The signal candidates for the $D_s^+\to\rho^+\phi,\phi\to\ee$ decay are selected by further requiring the invariant mass $M(\pi^+\pi^0)$ to be within $[0.60,0.95]$~GeV$/c^2$.
For the non-resonant $D_s^+\to\pi^+\pi^0\ee$ decay, we exclude events with $M(\ee)$ in the range of $[0.96,1.05]$~GeV$/c^2$ to reject potential LD effects from $D_s^+\to\rho^+\phi,\phi\to\ee$.
For the $D_s^+\to K^+\pi^0\ee$ and $D_s^+\to K_S^0\pi^+\ee$ decays, the contributions from the resonant $\phi\to\ee$ decay are expected to be insignificant with BFs of the order of $10^{-8}$ after taking the BF of $\mathcal{B}(\phi\to\ee)=(2.979\pm 0.033)\times 10^{-4}$~\cite{pdg} into account. 
Since this value is already beyond the sensitivity of \mbox{BESIII}, no further requirement on $M(\ee)$ is required for these two decays. After the application of all selection criteria, the signal efficiencies range from 5.3$\%$ to 25.1$\%$ for different decays, as outlined in Table.~\ref{tab:BF}. 
The variation in signal efficiencies is mainly caused by the different $M_{\rm rec}$ and $\Delta M$ requirements, and different tracking and PID efficiencies in the signal decays.

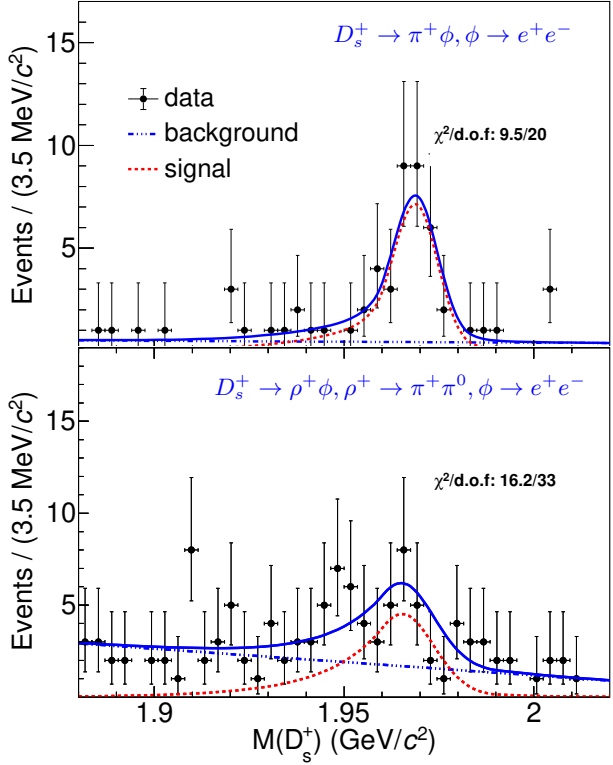
\begin{figure}[!htb]
\centering  
	\subfigure{
  \includegraphics[trim=0 66 0 2,clip,width=0.95\linewidth]{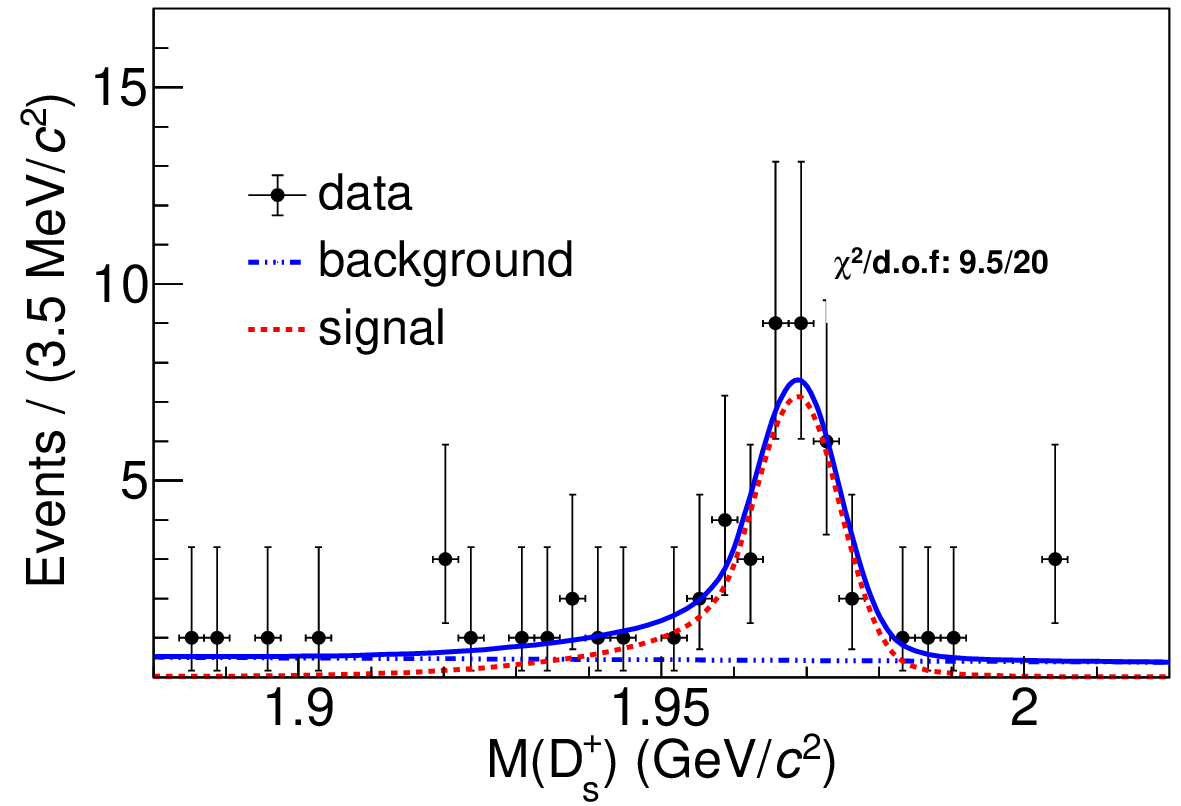}
  \begin{picture}(0,0)
    \put(-110,115){\color{blue}{$D_s^+\to\pi^+\phi,\phi\to\ee$}}
  \end{picture}
}
\vskip -6pt
	\subfigure{

  \includegraphics[trim=0 0 0 3,clip,width=0.95\linewidth]{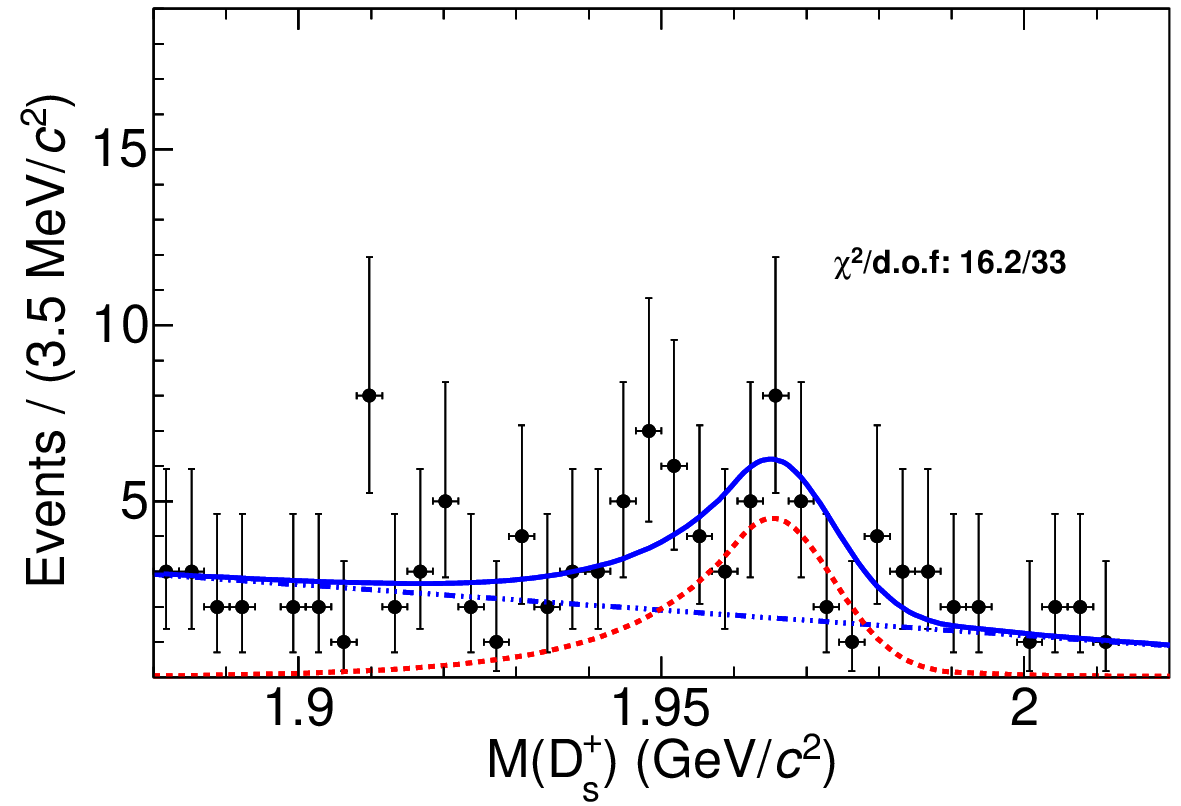}
  \begin{picture}(0,0)
    \put(-155,140){\color{blue}{$D_s^+\to\rho^+\phi,\rho^+\to\pi^+\pi^0,\phi\to\ee$}}
  \end{picture}
}
\caption{Fits to the $M(D_s^+)$ distributions for  $D_s^+\to\pi^+\phi,\phi\to\ee$ and $D_s^+\to\rho^+\phi,\rho^+\to\pi^+\pi^0,\phi\to\ee$. The signals are shown as the magenta dashed curves. The blue long-dashed curves are the combinatorial background components, and the dots with error bars are data. The $\chi^2/\rm{d.o.f}$ is displayed on each figure as an indication of the goodness of fit, where $\rm{d.o.f}$ is the number of degrees of freedom in each fit. The low $\chi^2/\rm{d.o.f}$ values are the result of the low statistics.}	
\label{fig:fit-R1}
\end{figure}

\begin{figure}[htp]
	\centering
  	\subfigure{
  \includegraphics[trim=0 62 0 0,clip,width=0.95\linewidth]{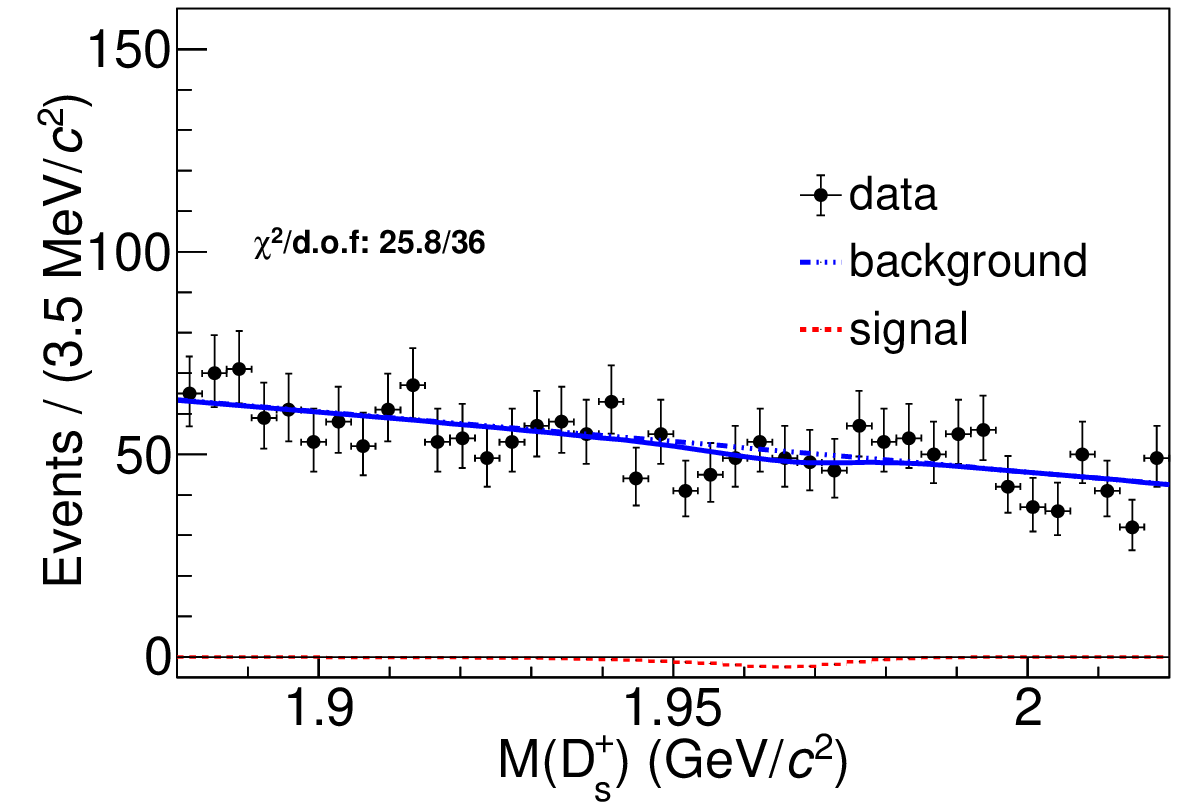}
  \begin{picture}(0,0)
    \put(-110,115){\color{blue}{$D_s^+\to\pi^+\pi^0\ee$}}
  \end{picture}
  }
  \vskip -6pt
  \subfigure{
  \includegraphics[trim=0 62 0 3,clip,width=0.95\linewidth]{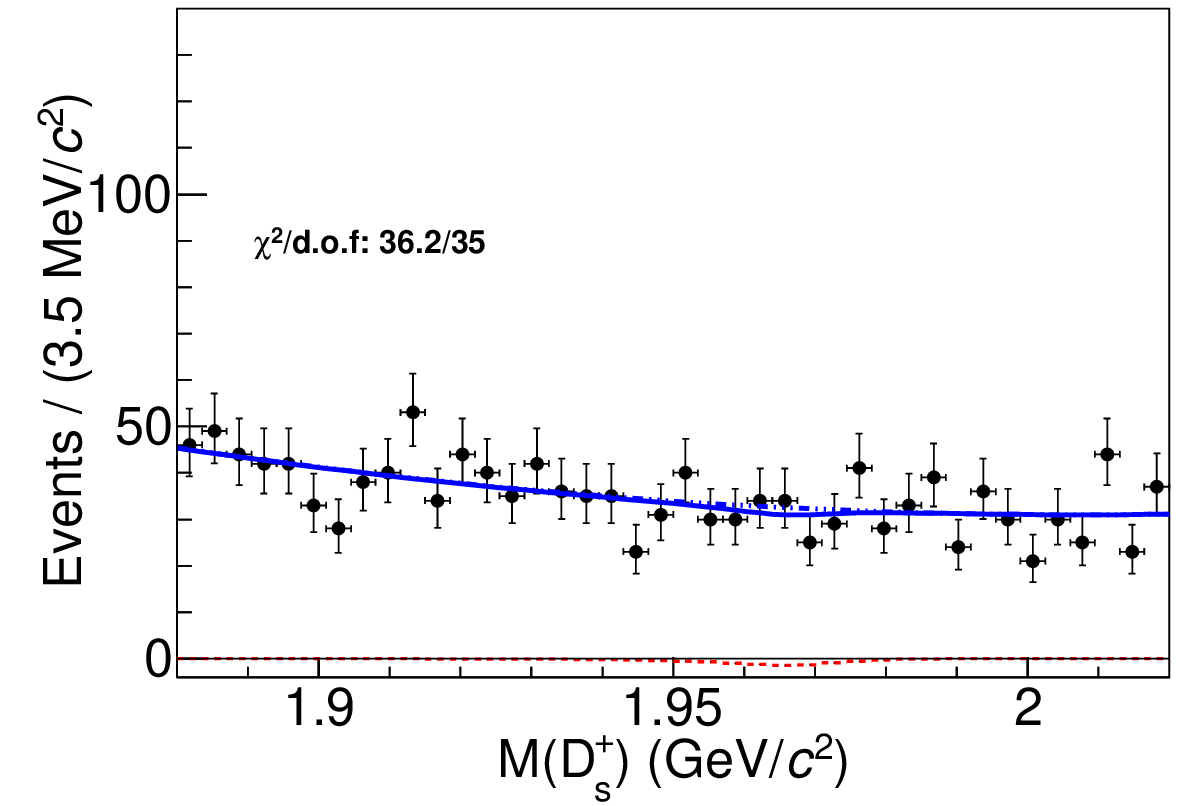}
  \begin{picture}(0,0)
    \put(-110,115){\color{blue}{$D_s^+\to K^+\pi^0\ee$}}
  \end{picture}   
  }
    \vskip -6pt
  \subfigure{
    \includegraphics[trim=0 0 0 3,clip,width=0.95\linewidth]{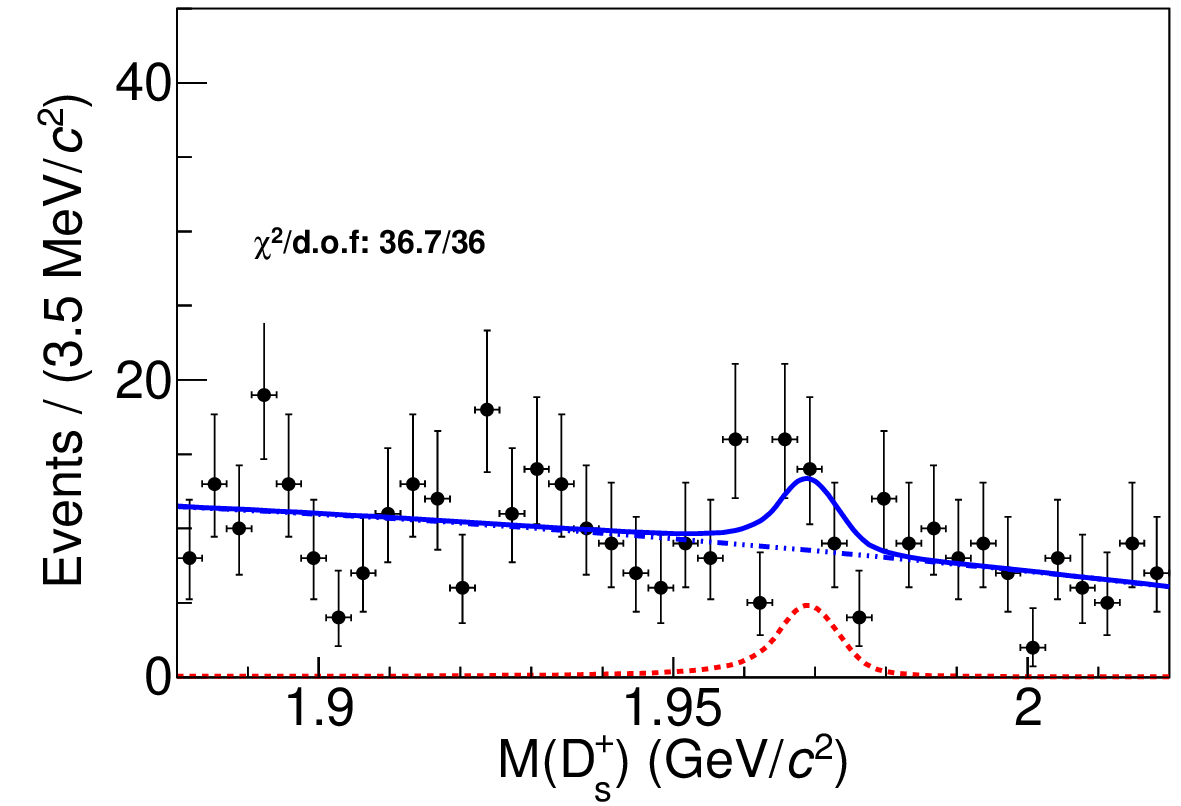}
  \begin{picture}(0,0)
    \put(-110,141){\color{blue}{$D_s^+\to K_S^0\pi^+\ee$}}
  \end{picture} 
	}
\caption{Fits to the $M(D_s^+)$ distributions for $D_s^+\to\pi^+\pi^0\ee$, $D_s^+\to K^+\pi^0\ee$, and $D_s^+\to K_S^0\pi^+\ee$. For $D_s^+\to\pi^+\pi^0\ee$, the $M(\ee)$ is required to be outside the $\phi$ mass window. The signals are shown as the magenta dashed curves.
The blue long-dashed curves are the combinatorial background components, and the dots
with error bars are data. The $\chi^2/\rm{d.o.f}$ is displayed on each figure to demonstrate the goodness of fit.}	
	\label{fig:fit-R2}
\end{figure}
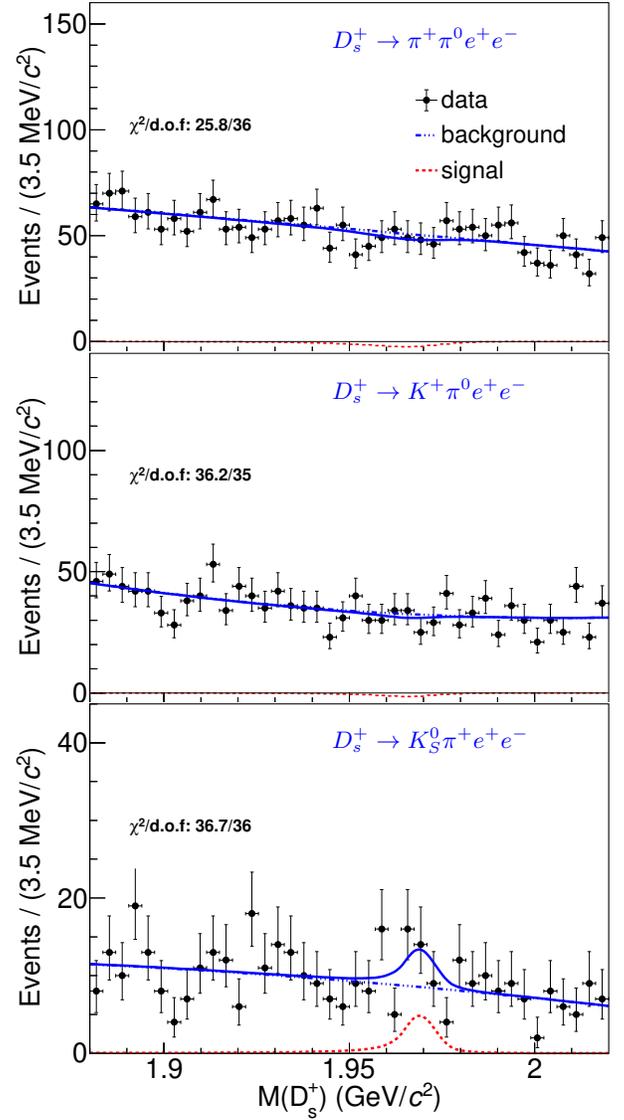

The signal yields of different $D_s^+$ decays in the data sample are determined independently using unbinned maximum likelihood fits to the invariant mass distributions of each final state in the range of 1.88 to 2.02~GeV$/c^2$. In these fits, the signal shapes are modelled by a double-sided Crystal Ball function~\cite{DSCB} plus a bifurcated Gaussian function (i.e., using asymmetric widths), with parameters determined from MC simulation. The residual backgrounds primarily originate from random combinations of the final states from $D_{(s)}$ meson decays and a small fraction of the continuum process $\ee\to q\bar{q}$, which vary smoothly across the fit range.
Consequently, the background shape is modelled by a first-order Chebychev polynomial function for $D_s^+\to\pi^+\phi,\phi\to\ee$ mode and $D_s^+\to\rho^+\phi,\phi\to\ee$ mode and a second-order Chebychev polynomial function for the others. The data distributions and fit results for two resonant decay modes are shown in Fig.~\ref{fig:fit-R1}. The obtained signal yields are $38.2_{-6.8}^{+7.8}$ for $D_s^+\to\pi^+\phi,\phi\to\ee$ and $37.8_{-9.6}^{+10.3}$ for $D_s^+\to\rho^+\phi,\phi\to\ee$, where the uncertainties are statistical only.
The signal from $D_s^+\to\pi^+\phi, \phi\to\ee$ is observed with a statistical significance of 7.8$\sigma$. The statistical significance is calculated by $\sqrt{-2\Delta\ln L}$, where $\Delta\ln L$ is the difference of the log-likelihoods with and without the signal component in the fit. Additionally, evidence for $D_s^+\to\rho^+\phi,\phi\to\ee$ is found for the first time with a statistical significance of 4.4$\sigma$. 
Figure~\ref{fig:fit-R2} shows the data distributions and fit results for the three four-body decay modes. No significant signal is observed, and the corresponding statistical significance is found to be less than 2$\sigma$ in each case.

The main sources of systematic uncertainties in the BF measurements include those associated with the total number of $ D_s^+D_s^{*-}$ pairs, the BFs of $\pi^0\to\gamma\gamma$, $K_S^0\to\pi^+\pi^-$ and $\rho^+\to\pi^+\pi^0$, the signal detection efficiency, and the signal extraction.
The systematic uncertainty from the total number of $D_s^+D_s^{*-}$ pairs is 0.4$\%$~\cite{ref:tot_DD}. An additional effect of possible contamination from $\ee\to D_s^+D_s^-$ and $\ee\to D_s^{*+}D_s^{*-}$ is estimated to be 0.3$\%$.
The uncertainties from the BFs of $\pi^0\to\gamma\gamma$, $K_S^0\to\pi^+\pi^-$, and $\rho^+\to\pi^+\pi^0$ are negligible~\cite{pdg}.
The uncertainties from the tracking (PID) efficiencies of $K^\pm$ and $\pi^\pm$ are estimated to be 0.8\% and 0.3\% (0.8\% and 0.5\%) per track, respectively~\cite{BESIII:2021dot}.
The uncertainties associated with the tracking and PID efficiencies of $e^{\pm}$ are studied with a radiative Bhabha ($\ee\to\gamma\ee$) control sample, which results in a systematic uncertainty of 1.0$\%$ for each of tracking and PID~\cite{BESIII:2019qci}.
The uncertainties related to the $\pi^0$ and $K_S^0$ reconstruction are assigned as 2.0$\%$ and 1.5$\%$, respectively~\cite{BESIII:2021dot,BESIII:2018mwk}.
The uncertainty due to the $\gamma$-conversion background veto is quoted as 1.8$\%$~\cite{BESIII:2015zpz}.

The uncertainty from the requirements of $\Delta M$ and $M_{\rm rec}$ is studied with a control sample of $D_s^{+}\to K^+K^-\pi^+$ candidates, giving a difference of 1.5$\%$ between data and MC simulation.
For the $D_s^+\to\rho^+\phi,\phi\to\ee$ decay, the uncertainty associated with the $\rho^+$ mass window requirement is estimated using the $D_s^+\to \rho^+\phi,\phi\to K^+K^-$ control sample. The resulting efficiency difference between data and MC simulation, 3.6$\%$, is assigned as the corresponding systematic uncertainty.
The signal MC samples of $D_s^+\to\pi^+\pi^0\ee$, $D_s^+\to K^+\pi^0\ee$, and $D_s^+\to K_S^0\pi^+\ee$ are generated uniformly in phase space. To estimate the uncertainty from the MC model, we generate a series of alternative MC samples by requiring the $\ee$ pairs to originate from a vector meson whose mass varied across the $M(\ee)$ spectrum.
The difference in efficiency between the two MC samples at each $M(\ee)$ bin are used to correct the phase space MC sample. The corresponding efficiency changes are assigned as mode-dependent systematic uncertainties:  3.7$\%$ for $D_s^+\to\pi^+\pi^0\ee$, 1.0$\%$ for $D_s^+\to K^+\pi^0\ee$, and 2.0$\%$ for $D_s^+\to K_S^+\pi^+\ee$. 
Moreover, additional MC samples are generated with the $\pi^+\pi^0, K\pi$ coming from $\rho, K^*$ decays and the resulting efficiency changes, compared to the phase-space MC samples, are assigned as systematic uncertainties:  9.3$\%$ for $D_s^+\to\pi^+\pi^0\ee$, 15.5$\%$ for $D_s^+\to K^+\pi^0\ee$, and 18.0$\%$ for $D_s^+\to K_S^0\pi^+\ee$.

The resolution difference between data and MC simulation for the signal shape is investigated using the signal from $D_s^+\to\pi^+\phi,\phi\to\ee$.  An alternative fit is performed with the signal modelled by the signal MC shape convolved with a Gaussian function, representing the resolution difference between data and MC simulation. The resulting signal yield changes by 0.8$\%$, which is assigned as the systematic uncertainty.
The systematic uncertainties associated with the background model are estimated by repeating the fit with the background MC simulated shape; the relative changes of the signal yields are 0.8$\%$ for  $D_s^+\to\pi^+\phi,\phi\to\ee$, and 4.7$\%$ for $D_s^+\to\rho^+\phi,\phi\to\ee$ decays.
For the upper limit measurements, the variation giving the largest upper limit on the signal yield is taken as the final result. 

Assuming that all the sources of systematic uncertainties (listed in the Supplemental Material~\cite{supp_mat}) are independent, the total systematic uncertainties are obtained to be 2.4$\%$ for $D_s^+\to\pi^+\phi,\phi\to\ee$, 6.6$\%$ for $D_s^+\to\rho^+\phi,\phi\to\ee$, 9.9$\%$ for $D_s^+\to\pi^+\pi^-\ee$, 15.9$\%$ for $D_s^+\to K^+\pi^0\ee$, and 18.3$\%$ for $D_s^+\to K_S^0\pi^+\ee$, by adding all sources in quadrature.

Using Eq.~\ref{eq:bf}, the measured BFs of the two LD measurements are
determined and presented in Table~\ref{tab:BF}, 
where the first uncertainty is statistical and the second systematic. 
These two values are consistent with the direct calculations of $\mathcal{B}(D_s^+\to\pi^+\phi)\cdot\mathcal{B}(\phi\to\ee)=(1.34\pm0.12)\times 10^{-5}$ and $\mathcal{B}(D_s^+\to\rho^+\phi)\cdot\mathcal{B}(\phi\to\ee)=(1.67\pm0.11)\times 10^{-5}$ from the PDG~\cite{pdg}.
For the three four-body decay modes, no significant signal is found. Therefore, we set upper limits on the BFs at the 90$\%$ confidence level.
Applying a Bayesian method~\cite{Zhu:2007zza}, a likelihood scan is performed by fixing the signal yield at various values. The effects of the systematic uncertainty are included by convolving the likelihood curve with a Gaussian function, where the standard deviation is set to the total systematic uncertainty. The corresponding normalized likelihood curves can be found in Ref.~\cite{supp_mat}.
These 90$\%$ upper limits for the four body decays are also listed in Table~\ref{tab:BF}. 

 \begin{table}[htb]
	\centering
	\caption{The signal yields ($N_{\rm{sig}}$), signal efficiencies, the BFs, and the 90$\%$ confidence level upper limits.}
	\label{tab:BF}
   \tabcolsep=0.15cm
   \renewcommand{\arraystretch}{1.3}
	\begin{tabular}{c    c  c c}
		\hline\hline
		\centering{Decay} & $N_{\rm{sig}}$ & $\epsilon$~($\%$) &  $\mathcal{B}~$($\times 10^{-5}$) \\ 
		\hline
    $D_s^+\to\pi^+\phi,\phi\to\ee$ & $38.2_{-6.8}^{+7.8}$ &  25.1 & $1.17^{+0.23}_{-0.21}\pm0.03$\\
  	$D_s^+\to\rho^+\phi,\phi\to\ee$ &$37.8_{-9.6}^{+10.3}$& 12.1  & $2.44^{+0.67}_{-0.62}\pm0.16$\\
   	$D_s^+\to\pi^+\pi^0\ee$ & ... & 7.4 & $<7.0$\\
    $D_s^+\to K^+\pi^0\ee$ & ... & 5.3 & $<7.1$\\
    $D_s^+\to K_S^0\pi^+\ee$ & ... &6.7 &$<8.1$\\
              \hline\hline
	\end{tabular}
 
\end{table}

In conclusion, we search for the rare decays $D_s^+\to h^+(h^0)\ee$ using a data sample corresponding to an integrated luminosity of $7.33$~fb$^{-1}$~\cite{BESIII:2020eyu} taken at $\ee$ c.m.~energies in the range $\sqrt{s}=4.128 - 4.226$~GeV. The $D_s^+\to\pi^+\phi,\phi\to\ee$ decay is observed with a statistical significance of 7.8$\sigma$.  Evidence of the $D_s^+\to \rho^+\phi,\phi\to\ee$ decay is found for the first time with a statistical significance of 4.4$\sigma$. The BFs of these two decays are measured to be
$\mathcal{B}(D_s^+\to\pi^+\phi,\phi\to\ee)=(1.17^{+0.23}_{-0.21}\pm0.03)\times 10^{-5}$ and 
 $\mathcal{B}(D_s^+\to\rho^+\phi, \phi\to\ee )=(2.44^{+0.67}_{-0.62}\pm0.16)\times 10^{-5}$,
where the first uncertainties are statistical and the second systematic. 
The BF of $D_s^+\to\pi^+\phi,\phi\to\ee$ is in agreement with the CLEO~\cite{CLEO:2010ksb} result quoted by the PDG~\cite{pdg}, with a precision improved by a factor of three. Our result is also in agreement with the BaBar result, $\mathcal{B}(D_s^+\to\pi^+\phi,\phi\to\ee)=(0.97\pm0.18)\times 10^{-5}$, obtained indirectly (based on the yields and efficiencies listed in Ref.~\cite{BaBar:2011ouc}) and thus not used by the PDG.
Both BFs are consistent with the products of the PDG values, $\mathcal{B}(D_s^+\to\pi^+\phi)\cdot\mathcal{B}(\phi\to\ee)$ and $\mathcal{B}(D_s^+\to\rho^+\phi)\cdot\mathcal{B}(\phi\to\ee)$~\cite{pdg} within uncertainties.
No significant signal of the four-body rare decays is observed, and the upper limits on the BFs of these decays are set to be $ \mathcal{B}(D_s^+\to\pi^+\pi^0\ee )<7.0\times 10^{-5}$, $\mathcal{B}(D_s^+\to K^+\pi^0\ee )<7.1\times 10^{-5}$, and $ \mathcal{B}(D_s^+\to K_S^0\pi^+e^+e^-)<8.1\times 10^{-5}$, at the 90$\%$ confidence level. 
These results represent the first upper limits on the BFs of these decays.
Our measurements improve our knowledge of LU in $c\to u l^+l^-$ and $D_s^+ \to V \gamma$ decays.

The BESIII Collaboration thanks the staff of BEPCII and the IHEP computing center for their strong support. This work is supported in part by National Key R$\&$D Program of China under Contracts Nos. 2020YFA0406400, 2020YFA0406300; National Natural Science Foundation of China (NSFC) under Contracts Nos. 11635010, 11735014, 11835012, 11935015, 11935016, 11935018, 11961141012, 12025502, 12035009, 12035013, 12061131003, 12192260, 12192261, 12192262, 12192263, 12192264, 12192265, 12221005, 12225509, 12235017; the Chinese Academy of Sciences (CAS) Large-Scale Scientific Facility Program; the CAS Center for Excellence in Particle Physics (CCEPP); Joint Large-Scale Scientific Facility Funds of the NSFC and CAS under Contract No. U1832207; CAS Key Research Program of Frontier Sciences under Contracts Nos. QYZDJ-SSW-SLH003, QYZDJ-SSW-SLH040; 100 Talents Program of CAS; The Institute of Nuclear and Particle Physics (INPAC) and Shanghai Key Laboratory for Particle Physics and Cosmology; European Union's Horizon 2020 research and innovation programme under Marie Sklodowska-Curie grant agreement under Contract No. 894790; German Research Foundation DFG under Contracts Nos. 455635585, Collaborative Research Center CRC 1044, FOR5327, GRK 2149; Istituto Nazionale di Fisica Nucleare, Italy; Ministry of Development of Turkey under Contract No. DPT2006K-120470; National Research Foundation of Korea under Contract No. NRF-2022R1A2C1092335; National Science and Technology fund of Mongolia; National Science Research and Innovation Fund (NSRF) via the Program Management Unit for Human Resources $\&$ Institutional Development, Research and Innovation of Thailand under Contract No. B16F640076; Polish National Science Centre under Contract No. 2019/35/O/ST2/02907; The Swedish Research Council; U. S. Department of Energy under Contract No. DE-FG02-05ER41374.

\end{document}